\documentclass[aps,prl,twocolumn,notitlepage,showpacs,floatfix,superscriptaddress]{revtex4-1}
\usepackage{dcolumn}
\usepackage{tabularx}
\usepackage{bm}
\usepackage{soul}
\usepackage{amsmath,amssymb,graphicx}
\usepackage[colorlinks=true,citecolor=blue,urlcolor=blue,linkcolor=blue]{hyperref}
\usepackage{environ}

\NewEnviron{eqnsplit}{%
\begin{equation}
\begin{split}
  \BODY
\end{split}
\end{equation}
}

\newcommand{\mrm}[1]{\mathrm{#1}}

\begin{document}
\title{Coupled spin-charge dynamics in magnetic van der Waals heterostructures}

\author{Avinash Rustagi}
\email{arustag@purdue.edu}
\affiliation{School of Electrical and Computer Engineering, Purdue University, West Lafayette, IN 47907}
\author{Abhishek Solanki}
\affiliation{School of Electrical and Computer Engineering, Purdue University, West Lafayette, IN 47907}
\author{Yaroslav Tserkovnyak}
\affiliation{Department of Physics and Astronomy, University of California, Los Angeles, CA 90095}
\author{Pramey Upadhyaya}
\email{prameyup@purdue.edu}
\affiliation{School of Electrical and Computer Engineering, Purdue University, West Lafayette, IN 47907}
\date{\today}
\begin{abstract}
We present a phenomenological theory for coupled spin-charge dynamics in magnetic van der Waals heterostructures. The system studied consists of a layered antiferromagnet inserted into a capacitive vdW heterostructure. It has been recently demonstrated that charge doping in such layered antiferromagnets can modulate the strength, and even the sign, of exchange coupling between the layer magnetizations. This provides a mechanism for electrically generating magnetization dynamics. The central result we predict here is that the magnetization dynamics reciprocally results in inducing charge dynamics. Such dynamics makes magnetic van der Waals heterostructures interesting candidates for spintronics applications. To this end, we also show that these systems can be used to convert sub-THz radiation induced magnetization dynamics into electrical signals. 
\end{abstract}

\maketitle
\textit{Introduction}| Two-dimensional (2D) magnetism offers opportunities to investigate phenomena driven by enhanced fluctuations, reduced symmetries, and nontrivial topology \cite{burch2018magnetism}. The recent discovery of stabilizing magnetic order in few layer systems held together by van der Waals forces (magnetic vdWs \cite{Novoselov2019_review2DMagnets,gong2017discovery}) have thus emerged as a promising avenue for fundamental research and technological applications of 2D magnets. Furthermore, by taking advantage of layer by layer assembly, magnetic vdWs can be interfaced with a wide range of materials to form heterostructures with desired functionality \cite{geim2013van}.

A functionality of fundamental interest is the ability to interconvert between spin and charge degrees of freedom. Such spin-charge conversion has been a key enabler of spintronics technology, which has attracted rigorous interest as an alternate to charge-based technology \cite{Zutic_Spintronics,han2018quantum}. Consequently, a number of fundamental phenomena allowing for conversion between spin and charge has been uncovered in the recent past, which includes reciprocal pairs such as spin torque-spin pumping \citep{Berger1996,slonczewski1996current,TserkovnyakSpumping}, spin Hall-inverse spin Hall effect \citep{dyakonov1971current,saitoh2006conversion,kimura2007RSHE,SinovaSHE}, Rashba Edelstein-inverse Rashba Edelstein effect \cite{edelstein1990spin,zhang2015charge,sanchez2013spin}, and direct-converse magneto-electric effect \cite{fiebig2005revival,zhang2019direct}. The search for alternate low-dissipation mechanisms and material platforms for efficient conversion between spin and charge is an active area of research \cite{han2018quantum}. 

Bilayer Chromium Iodide (CrI$_3$)-based vdW heterostructures have recently emerged as one such platform. CrI$_3$ is a vdW magnet with ferromagnetically ordered layers coupled via an antiferromagnetic interlayer coupling \cite{huang2017layer,sivadas2018stacking}. The interlayer coupling is sensitive to charge residing on each layer. This has been utilized to switch the ground state spin configuration in gated bilayer CrI$_3$ heterostructures between a layered antiferromagnetic and a ferromagnetic state electrically \cite{jiang2018controlling}. Reciprocity then dictates the existence of an inverse mechanism, wherein the charge distribution in the same heterostructure can be modulated by inducing magnetization dynamics. Furthermore, the large out-of-plane anisotropy, in combination with the antiferromagnetic coupling, gives rise to sub THz ferromagnetic resonance modes in the absence of external magnetic fields \cite{klein2018probing,lado2017origin}. This suggests the existence of a new reciprocal pair in gated bilayer CrI$_3$ heterostructures, which is capable of interconverting between spin and charge up to sub terahertz frequencies. In this Letter, we thus present a phenomenological theory of dynamical spin-charge coupling in vdW hetrostructures involving bilayer CrI$_3$. In particular, we predict the phenomena of charge pumping by magnetization dynamics, which can be utilized for electrical detection of antiferromagnetic resonance or converting GHz to sub THz radiation into an electrical signal.

\begin{figure}[!ht]
\centering
\includegraphics[width=0.49\textwidth]{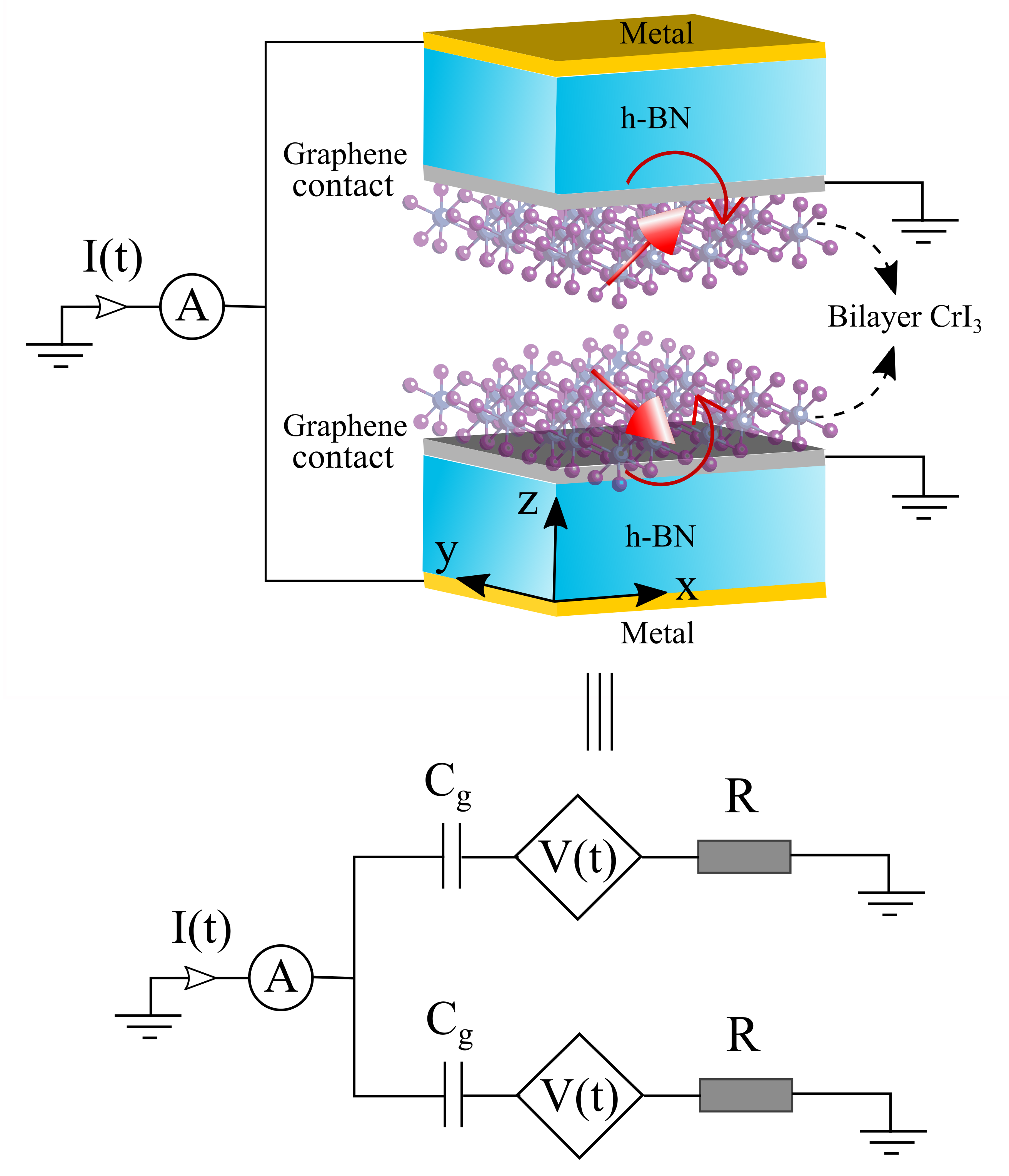}
\caption{\label{schematic_and_circuit}Schematic of the grounded (using graphene contacts) bilayer CrI$_3$ inserted into a h-BN capacitor for coupled spin-charge dynamics. Effective circuit model for the device structure accounting for the magnetization dynamics induced charge dynamics.}
\end{figure}

\textit{Structure and Model}| Motivated by recent experiments \cite{jiang2018controlling}, here we consider a bi-layer CrI$_3$ system inserted into a capacitor formed by hexagonal Boron-Nitride (h-BN), which is contacted by metal layers as shown in the left panel of Fig.~\ref{schematic_and_circuit}. The CrI$_3$ layers are themselves connected to ground (for example via contacting them by graphene layers). 
The intralayer magnetic moments in each layer of bi-layer CrI$_3$ are ferromagnetically aligned. However, the interlayer magnetic moments are antiferromagnetically coupled \cite{huang2017layer,klein2018probing}. Moreover, the magnetization of the layers have an out of plane easy axis, which arises due to anisotropic exchange interaction \cite{lado2017origin}. Thus, the free energy per unit area within the macrospin approximation in its minimal form consists of an easy axis uniaxial anisotropy and an antiferromagnetic interlayer exchange,
\begin{equation}
\mathcal{F} (\vec{m}_1,\vec{m}_2) = - K m_{1z}^2 - K m_{2z}^2 + J_\perp \, \vec{m}_1 \cdot \vec{m}_2.
\label{magnetic}
\end{equation}

The charges on the CrI$_3$ layers couple to the magnetic degrees of freedom giving rise to additional free energy $\mathcal{F}(\vec{m}_i, \sigma_i)$ terms. The functional form of $\mathcal{F}(\vec{m}_i, \sigma_i)$ is dictated by the time-reversal and structural symmetry of our van der Waals heterostructure. In particular, our structure has inversion symmetry under which the symmetric charge combination $ \sigma_1+\sigma_2$ remains invariant, while the anti-symmetric charge combination  $\sigma_1-\sigma_2$ changes sign. Noting that the magnetization $\vec{m}$ is not affected by structural inversion, the coupling between the magnetic and charge degrees of freedom to the lowest order in $\sigma_i$, and obeying structural inversion and time-reversal, can be written as \citep{ExtraTerms_FreeEnergy}: 
\begin{equation}
\mathcal{F}(\vec{m}_i, \sigma_i) = -\lambda \, (\sigma_1+\sigma_2) \vec{m}_1 \cdot \vec{m}_2.
\label{sc_coupl}
\end{equation}
This can be identified as doping-induced change in interlayer exchange coupling. We highlight that the experimental observation that asymmetric charge does not induce changes in the interlayer exchange \cite{jiang2018controlling} can be directly related to the structural symmetry of the van der Waals heterostructures. Additionally, the asymmetric charge can also couple to the magnetic degrees of freedom via a term of the form $\sim \zeta (\sigma_1-\sigma_2) (\vec{m}_1 - \vec{m}_2)\cdot \vec{H}$. This ``magneto-electric coupling"  has recently been observed in experiments \cite{huang2017layer}. However, the magnetoelectric effect is found to be less efficient for inducing magnetization dynamics when compared with the doping-induced interlayer exchange \cite{jiang2018controlling}. Thus, here the spin-charge coupling is restricted to the form given by Eq.~(\ref{sc_coupl}). 

Experimentally, the coupling given by Eq.~(\ref{sc_coupl}) has been utilized to electrically switch between a ferromagnetic and an antiferromagnetic configuration \cite{jiang2018controlling}. Reciprocally, a change in the magnetic configuration should alter charge density on the CrI$_3$ layer. We next derive an equivalent circuit for this coupled spin-charge dynamics. To this end, we need to supplement Eqs.~(\ref{magnetic}) and ~(\ref{sc_coupl}) with the electrical energy solely due to charges, which is given by:
\begin{equation}
\mathcal{F}(\sigma_1,\sigma_2) = \frac{\sigma_1^2}{2C_{\rm g}}+\frac{\sigma_2^2}{2C_{\rm g}}.
\label{electric}
\end{equation}
Here, $C_{\rm g}=\epsilon /d$ is the geometrical capacitance of the h-BN layer where $\epsilon$ and $d$ are the permittivity and thickness of h-BN, respectively \citep{Capacitor_Note}. 

Collecting the free energy terms involving charge densities, we thus have 
\begin{equation}
\mathcal{F}_{\sigma} = -\lambda (\sigma_1+\sigma_2) \vec{m}_1 \cdot \vec{m}_2 + \frac{\sigma_1^2}{2C_{\rm g}}+\frac{\sigma_2^2}{2C_{\rm g}}.
\end{equation}
Assuming $R$ to be the series combination of the resistance of the external circuit and the resistance between graphene contacts and the CrI$_3$ layer, we can draw an effective circuit for the van der Waals heterostructure as shown in Fig.~\ref{schematic_and_circuit}. The central result of this model is that a dynamic magnetization gives rise to a time-dependent voltage
\begin{equation}
V = -\partial \mathcal{F}(\vec{m}_i, \sigma) / \partial \sigma = \lambda \, \vec{m}_1 \cdot \vec{m}_2.
\label{volt_eq} 
\end{equation}
This circuit is the first main result of this letter.

\begin{figure}
\centering
\includegraphics[width=0.49\textwidth]{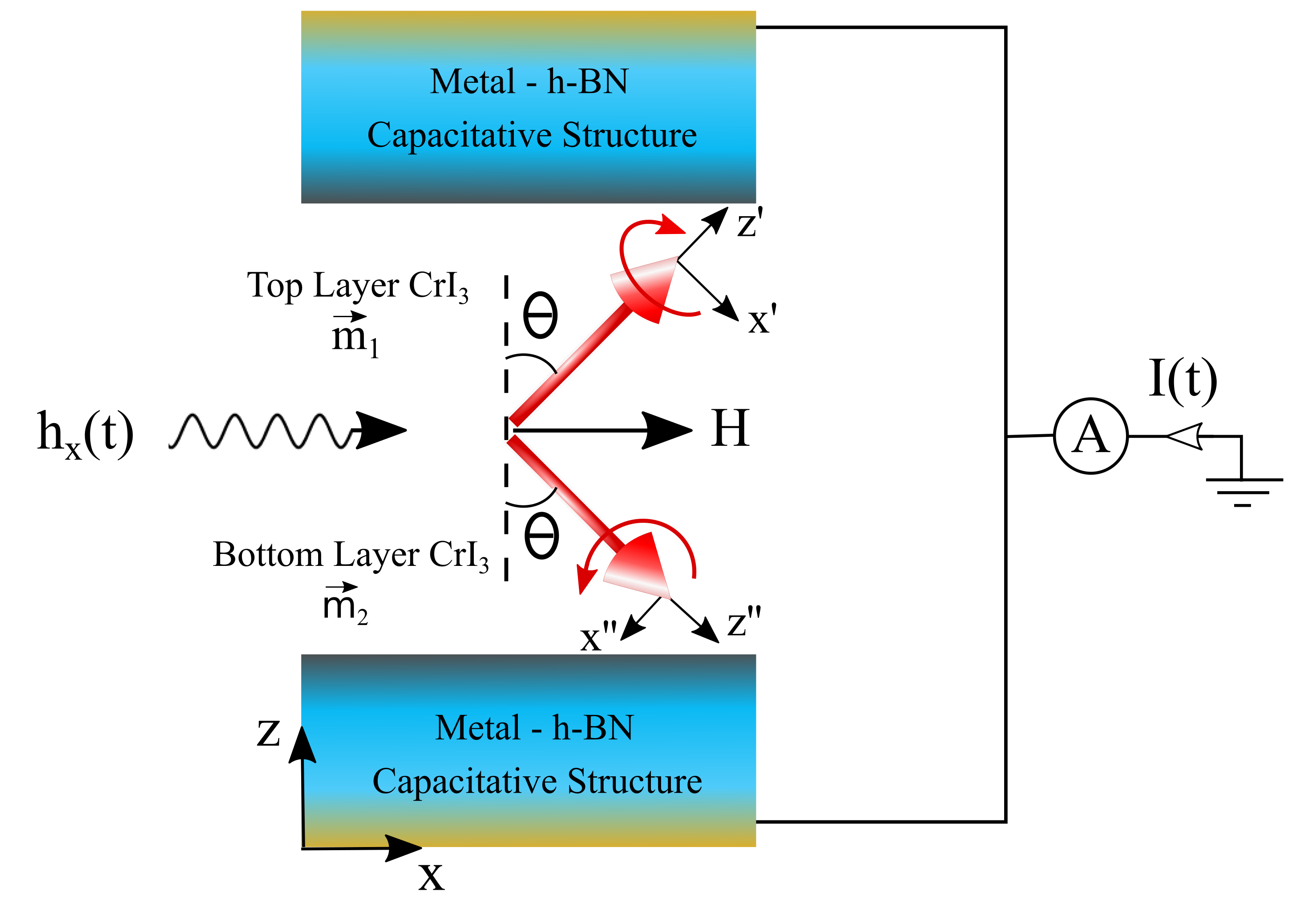}
\caption{\label{AFMR_setup} Schematic of the proposed AFMR setup. The equilibrium magnetizations cant by an angle $\theta$ due to the applied dc magnetic field $H$ and the AFMR mode is excited by the radiation field $h_x(t)$. }
\end{figure}

\textit{Charge pumping}| To demonstrate the effects of the reciprocal process of magnetization dynamics induced voltage, we propose to utilize the antiferromagnetic resonance (AFMR) set up (see Fig.~\ref{AFMR_setup}). In this case, the antiferromagnet is subjected to an electromagnetic radiation in the presence of a dc magnetic field. A resonant excitation of the magnetic order parameter occurs when the frequency of the electromagnetic radiation coincides with the frequency of the inherent magnetic dynamical modes of the antiferromagnet (which can be tuned by varying the strength of the dc magnetic field). The second main result of this letter is that in the CrI$_3$-based vdW heterostructures the absorbed radiation also gives rise to an electrical signal. In particular, when the RC time associated with vdW heterostructure is smaller than the timescale associated with magnetization dynamics \citep{RCTime_Note}, the induced voltage results in a flow of charge current in the external circuit, which is given by:
\begin{equation}
\label{current}
j_{ext}(t) = \dot{\sigma}_{1}+\dot{\sigma}_{2} = 2 \lambda\, C_{\rm g} \,\partial_t \vec{m}_1 \cdot \vec{m}_2.
\end{equation}
We next calculate analytically the charge current pumped due to AFMR within the linear response, which is corroborated by numerical simulations.

\begin{figure*}
\centering
\includegraphics[width=0.99\textwidth]{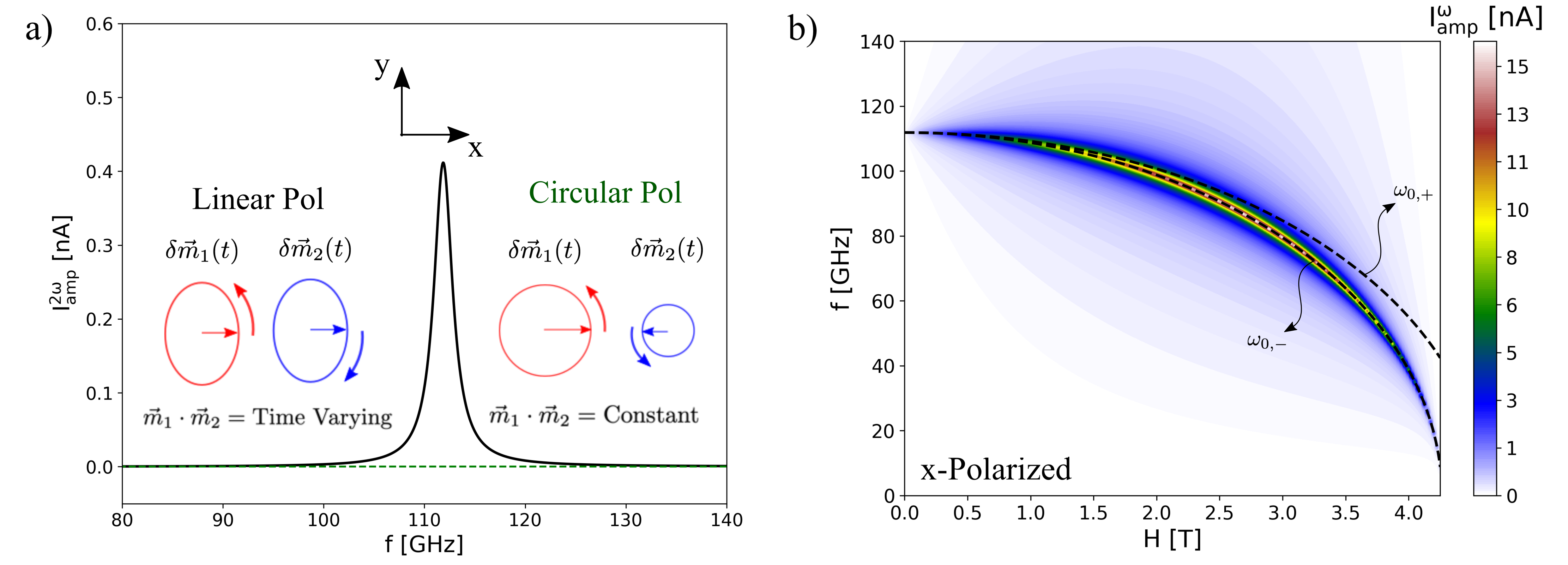}
\caption{\label{CurrentPlot} a) Electrical response ($I^{2\omega}_{\mathrm{amp}}$) from Kittel modes at $2\omega$ from magnetization dynamics induced by linearly (solid) and circularly (dashed) polarized radiation fields as a function of excitation field frequency in absence of dc field $H=0$. The inset displays the precession of magnetization deviations transverse to the equilibrium orientation, of the top and bottom CrI$_3$ layers for both polarizations. b) Evaluated current amplitude at excitation field frequency $I^\omega_{\mathrm{amp}}$ as a function of excitation magnetic field frequency and the in-plane dc magnetic field for x-polarized excitation field. Dashed lines show the two AFMR modes (Eq.~(\ref{AMFRmodes})) as a function of in-plane dc field.}
\end{figure*}

The magnetization dynamics are governed by the Landau-Lifshitz-Gilbert (LLG) equation,
\begin{equation}
\label{LLG_equation}
\dot{\vec{m}}_{i} = -\gamma\, \vec{m}_{i} \times \left[ \vec{H}_{\text{eff},i} + \vec{H}_{\text{ext}} + \vec{h}_\sim \right] + \alpha \, \vec{m}_{i} \times \dot{\vec{m}}_{i}
\end{equation}
where index $i=1,2$ correspond to the two layers, $\vec{m}_i$ is the unit magnetization, $\gamma>0$ is the gyromagnetic ratio, $\alpha$ is the damping, $\vec{H}_{\text{eff},i} = -\partial_{\vec{m}_i} \mathcal{F}$, and $\vec{h}_\sim$ represents the ac excitation field. We take the dc external field to be oriented along the $x$ axis, that is $\vec{H}_{\text{ext}} = H \hat{x}$. This dc field cants the otherwise $z$ axis oriented ${\vec{m}}_{i}$ to align with $z^{\prime}$, which makes a polar angle $\theta$ with the $z$ axis (see Fig.~\ref{AFMR_setup}). The polar angle satisfies $\sin\theta = H/(H_k+2H_J)$ when $H<H_k+2H_J$ and  $\theta=\pi/2$ for $H>H_k+2H_J$. Here $H_k = 2K/M_s$ is the anisotropy field and $H_J = J_\perp/M_s$ is the interlayer exchange field. The eigenmodes, found by solving the LLG linearized about the tilted equilibrium \cite{von2016ferromagnetic}, correspond to the antiferromagnetic resonance (AFMR) frequencies given by
\begin{equation}
\omega_\mrm{0,\pm} = \sqrt{[\omega_{eq}\pm\omega_J][\omega_{eq}\mp \omega_J \cos2\theta - \omega_k \sin^2\theta]} 
\label{AMFRmodes}
\end{equation}
where $\omega_{k} = \gamma H_{k}$, $\omega_{J} = \gamma H_{J}$, and $\omega_{eq} = \gamma H_{eq}$ are the frequencies corresponding to the anisotropy, interlayer exchange, and equilibrium fields. The equilibrium field experienced by the magnetizations has contributions from external, anisotropy, and interlayer exchange fields $H_{eq} = H \sin\theta + H_k \cos^2\theta + H_J \cos2\theta$. The large easy axis out of plane anisotropy in few layer CrI$_3$, corresponding to a field of $H_k = 3.75 \text{ T}$, results in a large spin wave gap $\sim 110 \text{ GHz}$ \citep{lado2017origin} for $\theta=0$. The presence of the in-plane dc field $H$ lowers the AFMR frequency by effectively offsetting the out of plane anisotropy field. Also, the two degenerate Kittel modes in absence of the dc field, are now non-degenerate [c.f. Fig.~\ref{CurrentPlot}].

To calculate the charge current pumped by above modes, we need to find $\vec{m}_1 \cdot \vec{m}_2$ [ c.f. Eq.~(\ref{current})]. For this purpose, we also evaluate the deviations transverse to the tilted equilibrium magnetizations ($z'$ and $z''$ axis), $(\delta m_{1x}',\delta m_{1y}')$ and $(\delta m_{2x}'',\delta m_{2y}'')$, within the linearized LLG. A convenient choice for the solution is in terms of the collective coordinates $X_{\pm}=\delta m_{1x}^{\prime} \pm \delta m_{2x}^{\prime \prime}$ and $Y_{\pm}=\delta m_{1y}^{\prime} \pm \delta m_{2y}^{\prime \prime}$, which gives : 
\begin{equation}
\label{mdotP}
\begin{split}
\vec{m}_1 \cdot \vec{m}_2 
=  \dfrac{1}{4} \bigg[ -\cos2\theta \, X_+^2 + Y_+^2 - 4 \cos2\theta + \\
\qquad \cos2\theta\, X_-^2- Y_-^2 + 2\sin2\theta\,X_- \bigg], 
\end{split}
\end{equation}
where
\begin{equation}
\label{linearized_m}
\begin{split}
\left(\begin{array}{c}
X_+ \\ 
Y_+
\end{array} \right) &= \text{Re}\left[ \dfrac{2\gamma\, h_y e^{-i\omega t}}{\omega^2 - \omega_{0,+}^2 + i \omega \Delta\omega_{+}} \left(\begin{array}{c}
i\omega \\ 
i\alpha\omega- \omega_{2+}
\end{array} \right) \right] \\
\left(\begin{array}{c}
X_- \\ 
Y_-
\end{array} \right) &= \text{Re}\left[ \dfrac{2\gamma h_x \cos\theta e^{-i\omega t}}{\omega^2 - \omega_{0,-}^2+ i \omega \Delta\omega_{-}} \left(\begin{array}{c}
i\alpha\omega- \omega_{1-} \\ 
-i \omega
\end{array} \right) \right]
\end{split}
\end{equation}
assuming $\alpha \ll 1$. Here, $h_x$ and $h_y$ are the $x$ and $y$ components of the oscillating field $\vec{h}_\sim$, $\omega_{1 \pm}= \omega_{eq} \pm \omega_J$, $\omega_{2 \pm}= \omega_{eq} \mp \omega_J \cos2\theta - \omega_k \sin^2\theta $, and $\Delta \omega_\pm = \alpha (\omega_{1\pm}+\omega_{2\pm})$. 
An important inference to make from Eq.~(\ref{mdotP}) is that the magnetization dot product has both quadratic and linear terms in the collective coordinates. This implies that the charge dynamics can have response at both $\omega$ and $2\omega$, where $\omega$ is the excitation field frequency. We next analyze key experimental signatures of the proposed charge pumping. 

\textit{Charge dynamics signatures}| We begin by discussing the charge pumping in the absence of the external dc field $H=0$ when the canting angle $\theta=0$. In Fig.~\ref{CurrentPlot}(a), we plot the amplitude of the charge current pumped as a function of the excitation frequency of external radiation, which is obtained by substituting the numerical solution of Eq.~(\ref{LLG_equation}) into Eq.~(\ref{current}). Here, the AFMR are the well-known Kittel modes $\omega_\mrm{Kittel} = \gamma \sqrt{H_k(H_k+2H_J)}$\citep{kittel1952AFMR}, which could be excited via a circular or linearly polarized radiation. We emphasize first that when exposed to circularly polarized oscillatory fields, i.e. $\vec{h}_\sim= h (\hat{x} \pm i \hat{y})/\sqrt{2}$, the eigenmodes do not lead to any charge dynamics (i.e. $\partial_t \vec{m}_1 \cdot \vec{m}_2 = 0$). This is because in this case the magnetizations of the two layers precess keeping the angle between the magnetizations constant. On the other hand, in the presence of linearly polarized excitation fields ($x$ and $y$-polarized), there is observable charge dynamics. Furthermore, from Eq.~(\ref{mdotP}) we see that in the absence of external field (i.e. when $\theta$ = 0), only quadratic terms in the magnetization dot product exists. Thus, the lowest order charge dynamics arising from the magnetization dynamics for $H=0$ has a response at twice the excitation frequency ($I^{2\omega}_{\mathrm{amp}}$) which at resonance ($\omega = \omega_{0,-}$) is approximately given by
\begin{equation}
\label{IAmp_zerotheta}
I^{2\omega}_{\mathrm{amp}} \approx 2 \lambda A C_{\rm g} \dfrac{\gamma^2 h_x^2 (\omega_{1-}^2+\omega_{0,-}^2)}{ \omega_{0,-} \Delta \omega_{-}^2}
\end{equation} 
which is in agreement with the amplitude evaluated from the numerical solution of Eq.~(\ref{LLG_equation}) plotted in Fig.~\ref{CurrentPlot}(a).

A larger current response is generated in the presence of a DC canting field. This can be seen from Eq.~(\ref{mdotP}), where due to non-zero $\theta$ we obtain a contribution to the charge current of the form $\sim 2\sin2\theta\,\partial_t \,X_-$. This contribution, being linear in deviation, pumps a larger charge current (when compared to the $H=0$ case) oscillating at the frequency of incident radiation. The on-resonance ($\omega = \omega_{0,-}$) amplitude of charge current can be approximated from Eq.~(\ref{mdotP}) and Eq.~(\ref{linearized_m}), given by
\begin{equation}
\label{IAmp_theta}
I^{\omega}_{\mathrm{amp}} \approx 2 \lambda A C_{\rm g} \sin2\theta \cos\theta \dfrac{\gamma h_x \sqrt{\omega_{1-}^2+\alpha^2 \omega_{0,-}^2}}{ \Delta \omega_{-}}.
\end{equation} 

In Fig.~\ref{CurrentPlot}(b), we plot the amplitude of this current ($I^{\omega}_{\mathrm{amp}}$) as a function of in-plane dc magnetic field and excitation frequency for an $x-$ polarized oscillating magnetic field evaluated from substituting the numerical solution of Eq.~(\ref{LLG_equation}) into Eq.~(\ref{current}). The linear response estimate in Eq.~(\ref{IAmp_theta}) agrees with the numerical result. The linear term contribution $\sin2\theta\,\partial_t \,X_- \sim \omega \sin2\theta\,X_-$ peaks at an intermediate canting angle. This is because while the AFMR frequency decreases with the in-plane DC field, the deviation in magnetization increases as the effective field seen by the magnetization gets reduced (since the DC field offsets the easy axis anisotropy). The competition between these two leads to an increase in the current amplitude at intermediate DC field strengths. For typical experimentally realized values of the parameters (see Table.~\ref{parameters_CrI3}), this peak current gives a value of $\sim$ 1.5 nA per Oe of oscillating magnetic field, well within the reach of  experiments.
\begin{table}
\caption{\label{parameters_CrI3} CrI$_3$ Parameters used in calculations.}
\begin{ruledtabular}
\begin{tabular}{lr}
Parameter & Value \\
\hline
Saturation Magnetization $M_s$ \cite{lado2017origin} & 1.37 $\times 10^{-5}$ emu/cm$^2$\\
Easy axis anisotropy $K$ \cite{lado2017origin} & 0.2557 erg/cm$^2$ \\
Interlayer Exchange $J_\perp$ \cite{soriano2019interplay} & 0.0354 erg/cm$^2$ \\
Spin-Charge Coupling $\lambda$ \cite{jiang2018controlling} & 1 mV\\
Area of bilayer CrI$_3$ $A$ & 1 $\rm \mu$m$^2$
\end{tabular}
\end{ruledtabular}
\end{table}

\textit{Conclusions and outlook}| In this letter, we constructed a phenomenological theory for coupled spin-charge dynamics in symmetrically gated vdW heterostructures of bilayer CrI$_3$. We find that the spin-charge coupling can be classified into doping-induced modifications of magnetic properties and the so-called magneto-electric coupling, which represents the interaction of electric field with the difference of magnetizations in each layer. The structural symmetries restrict the form of former (latter) to be directly proportional to the symmetric (antisymmetric) combination of charge doping within each layer. Motivated by the experimental observation of large  doping-induced changes in interlayer exchange coupling, we specifically construct an effective circuit theory for coupled spin charge dynamics including this term, which if needed can similarly be extended to include other spin charge couplings. 

A central finding of this theory is that the magnetization dynamics induces a voltage, which is proportional to the dot product of magnetizations within each layer. As an experimental signature of this effect, we proposed and evaluated the charge current pumped by magnetization dynamics induced by the absorption of electromagnetic radiation. This can be utilized for probing magnetic excitations electrically, in addition to tunneling magnetoresistance \cite{klein2018probing}, and harvesting electromagnetic radiation via conversion into an electrical signal with following features. In the absence of any dc magnetic field, the resulting charge dynamics have a response at twice the excitation field frequency when exposed to linearly polarized time-dependent electromagnetic fields. These high frequency AFMR modes can be softened by applying a dc magnetic field, which additionally allows a first harmonic response when time-dependent electromagnetic field is polarized along the dc field direction. 

Future works should explore the implications and opportunities offered by  the reciprocity dictated pair of `voltage $\leftrightarrow$ magnetization dynamics' in addition to the proposed AFMR setup. For example, thanks to the easy integration of various vdW materials, spin transistors have recently been fabricated \cite{jiang2019spin}. Here, a gate voltage-dependent magnetic configuration of bilayer CrI$_3$ controls the flow of tunneling charge current due to the tunneling magnetoresistance and through shift in the chemical potential of the vdW layers. Solving coupled spin-charge circuits in such vdW magnet-based spin transistors, will be addressed elsewhere. 

We thank Vaibhav Ostwal, Se Kwon Kim, and Kin Fai Mak for helpful discussions. A.R., A.S., and P.U. acknowledge support from the National Science Foundation through Grant No. DMR-1838513. Collaboration with Y.T. was supported by the National Science Foundation through Grant No. ECCS-1810494.

\bibliography{refs}

\begin{thebibliography}{33}
\bibitem [{Ext()}]{ExtraTerms_FreeEnergy}%
  \BibitemOpen
  \href@noop {} {}\bibinfo {note} {In addition, terms of the form $\sim
  (\sigma_1+\sigma_2) m_{i,z}^2$ i.e. doping-induced change in uniaxial
  ansiotropy, are also allowed. However, motivated by the experimental
  observation that the percentage change in anisotropy is typically smaller
  than that in the exchange interaction, i.e. $\Delta K/K \ll \Delta J/J$
  \cite{jiang2018controlling}, we have not included those terms
  here.}\BibitemShut {Stop}%
\end{thebibliography}
\end{document}